\documentstyle{mn}

\input{epsf.sty}

\begin{document}

\title{Outbursts of Irradiated Accretion Discs}
\author[A. R. King]{A. R. King\\
Astronomy Group, University of Leicester,
Leicester, LE1~7RH
}

\newcommand{\lta}{{\small\raisebox{-0.6ex}{$\,\stackrel
{\raisebox{-.2ex}{$\textstyle <$}}{\sim}\,$}}}
\newcommand{\gta}{{\small\raisebox{-0.6ex}{$\,\stackrel
{\raisebox{-.2ex}{$\textstyle >$}}{\sim}\,$}}}   

\maketitle
\begin{abstract}
I solve analytically the viscous evolution of a
self--irradiated accretion disc, as seen
during outbursts of soft X--ray transients.
The solutions predict steep power--law X--ray decays $L_X \sim
(1+ t/t_{\rm visc})^{-4}$, changing to $L_X \sim (1-t/t'_{\rm
visc})^4$ at late times, where $t_{\rm visc}, t'_{\rm visc}$ are
viscous timescales. These forms
closely resemble the approximate exponential and linear decays 
inferred by King and Ritter (1997) in these two regimes.
The decays are much steeper than for unirradiated discs because the
viscosity is a function of the central accretion rate rather than of
local conditions in the disc.

\end{abstract}

\section{INTRODUCTION}
\label{sec:intro}
Soft X--ray transients (SXTs) are close binary systems with black--hole or
neutron--star primaries; they undergo large outbursts, reaching X--ray
luminosities of order the Eddington limit. The outbursts have some
resemblance to those of dwarf novae, in which the primary is a white
dwarf, with the important difference that the timescales are far
longer. SXT outbursts last of order a year, and recur at intervals of
several years or greater, whereas these timescales are typically days
and weeks for dwarf novae. It is by now generally accepted that dwarf
nova outbursts result from a thermal--viscous disc instability (see
e.g. Cannizzo 1993 for a review). While this picture naturally
explains observed dwarf nova timescales it requires essentially ad hoc
modification to explain SXT light curves. However a crucial observed difference
between dwarf nova disc and those in SXTs is that the latter are
heavily irradiated by the central X--ray source (van Paradijs \&
McClintock, 1994; Shahbaz \& Kuulkers, 1997).
A recent paper (King \& Ritter, 1997, hereafter KR) has pointed out
that this fact
inevitably lengthens the entire SXT outburst cycle, as irradiation
will prevent the encroachment of the cooling front which terminates
a dwarf nova outburst. Instead, the disc can only return to the cool
state as its central accretion rate $\dot M_c$ 
is reduced, which in turn requires
most of the disc mass to be accreted. KR modelled this effect by
simple dimensional relations and showed that $\dot M_c$ (and thus the
X--ray emission) will decay roughly exponentially on the hot--state
viscous time of the irradiated disc. This prediction, and the related
timescales and luminosities, are in good agreement with observed SXT
light curves (see Tanaka \& Shibazaki, 1996 and Chen, Shrader \&
Livio, 1997 for recent reviews).

KR pointed out that long decays
were a feature of the now--discarded mass transfer burst model of
dwarf novae, which shared the property that the outburst could only
terminate through the accretion of most of the disc mass. However
Cannizzo (1997) has argued that since the mass transfer burst model
gives rather shallow power--law decays (typically $\dot M_c \sim
t^{-(1+\epsilon)}$ with $\epsilon << 1$; Pringle, 1974; Cannizzo, Lee
\& Goodman, 1990) rather than exponentials, the irradiated--disc model
of KR would give a poor representation of observed SXT light curves. I
consider this question further in this paper. 

In the unirradiated disc models considered by Pringle (1974) and 
Cannizzo, Lee \& Goodman, (1990) 
the viscosity controlling the time evolution of the disc is a function
of purely {\it local} variables, either through assumed power--law
dependences (Pringle) or the alpha--prescription 
\begin{equation}
\nu = \alpha c_SH \label{eq1}
\end{equation}
with the sound speed $c_S \propto T_m^{1/2}$ and scaleheight
\begin{equation}
H = c_S\biggl({R^3\over GM}\biggr)^{1/2}  \label{eq2}
\end{equation}
fixed by the disc midplane temperature $T_m$ resulting from local viscous
dissipation (here $R$ is the radial disc coordinate and $M$ the
central mass). The irradiated--disc model of KR is fundamentally
different: the temperature and viscosity are
no longer fixed locally, but instead by $\dot M_c$, and have 
explicit time dependences. I shall find analytic solutions of the
disc diffusion equation for this case. These give much
steeper power--law behaviour for $\dot M_c$, which is in practice
indistinguishable from the exponential decay predicted by KR.

\section{IRRADIATED DISCS}
The surface temperature of a disc irradiated by a central source is given by
\begin{equation}
T_{\rm irr}(R)^4 = {\eta \dot M_{\rm c}c^2(1-\beta)\over 4\pi \sigma R^2}
\biggl({H \over R}\biggr)^n
\biggl[{{\rm d}\ln H\over {\rm d}\ln R} - 1\biggr], 
\label{eq3}
\end{equation}
where $\eta$ is the efficiency of rest--mass energy conversion into X--ray 
heating, $\dot M_{\rm c}$ the central accretion rate,
$H$ the disc scaleheight at disc radius $R$, $\beta$ 
the albedo of the disc faces, and the factor in square 
brackets lies between 1/8 and 2/7. The index $n = 1, 2$ for 
irradiation by a central point source or the inner disc respectively
(van  Paradijs, 1996; Fukue, 1992; King, Kolb \&
Szuszkiewicz 1997). Neutron--star accretion always has $n=1$, while
$n=2$ is possible in the black hole case
because of the lack of a hard stellar surface; this effect probably
allows black--hole binaries to be transient at mass transfer rates which
would otherwise give a persistent system (King, Kolb \&
Szuszkiewicz 1997). However during outbursts, observations of a strong
power--law continuum strongly suggest that the
disc develops an extended X--ray corona, thereby reverting to $n=1$. The
distinction between these two possibilities will in any case
be unimportant for the
purposes of this paper (I shall assume $n=1$ below): 
the main effect comes from the radial dependence of
$T_{\rm irr}$, which is the same regardless of $n$. The ratio $H/R$ is
roughly constant 
in a disc, so $T_{\rm irr}$ falls off as $R^{-1/2}$. Thus for a 
disc with a large enough ratio of outer radius $R_0$ to inner radius
$R_*$, $T_{\rm irr}$ dominates
the disc's own effective temperature $T_{\rm eff}$, given by
\begin{equation}
T_{\rm eff}^4 = {3GM\dot M\over 8\pi\sigma R^3}
\label{eq3a}
\end{equation}
(e.g. Frank et al., 1992)
which goes as $R^{-3/4}$ (here $\dot M$ is the accretion rate at
radius $R$). Thus 
\begin{equation}
{T_{\rm irr}^4\over T_{\rm eff}^4} = 
       {4\over 3}{R\over R_s}{\dot M_{\rm c}\over \dot M}\eta(1 - \beta)
       {H \over R}
\biggl[{{\rm d}\ln H\over {\rm d}\ln R} - 1\biggr],
\label{eq3b}
\end{equation}
where $R_s = 2GM/c^2$ is the Schwarzschild radius of the central
object. In X--ray binaries one typically finds $H/R \simeq 0.2, \beta
= 0.9$ (de Jong, van Paradijs \& Augusteijn; 1996), while the last
factor on the rhs lies between 2/7 and 1/8. Further, the inner disc
radius $R_*$ is close to $R_s$ in all X--ray binaries, so the
efficiency $\eta$ is $\gta 0.1$. With a typical value $\eta =
0.2$ we find
\begin{equation}
{T_{\rm irr}^4\over T_{\rm eff}^4} \simeq 6.7\times 10^{-4}{R\over R_s}
\label{eq3c}
\end{equation}
for a quasi--steady disc ($\dot M \simeq \dot M_{\rm c}$). Since we
typically have $R_s \lta 2\times 10^6$ cm and a disc
radius $R_0 > 10^{11}$ cm, we find $T_{\rm irr}^4/T_{\rm eff}^4 \gta 34$. 

This result shows that irradiation dominates the output of the disc over
most of its surface in any situation where the central luminosity is
significant. In fact we can go further and show that irradiation
completely alters the disc structure over most of its area. For
$T_{\rm irr}$ generally exceeds the
midplane temperature $T_m$ of the outer disc, given by radiative equilibrium
as $T_m^4 \sim \tau T_{\rm eff}^4 \sim 10T_{\rm eff}^4$. Here 
$\tau \sim 10$ is the vertical optical depth of the unirradiated (cool)
outer disc, with typical radii $R_0 \gta 10^{11}$ cm (e.g. Frank
et al., 1992). The outer disc therefore develops an almost
isothermal vertical structure, and we can replace $T_m$ in $c_S$ by 
\begin{equation}
T_{\rm irr} \propto \dot M_c^{1/4}R^{-1/2} \label{eq4a}
\end{equation}
so that (\ref{eq1}, \ref{eq2}) give 
\begin{equation}
\nu = \nu_*\biggl({\dot M_c\over \dot M_*}\biggr)^{1/4}
             \biggl({R\over R_*}\biggr), 
\label{eq4}
\end{equation}
where $\nu_*, \dot M_*$ are the viscosity and accretion rate at the
inner disc radius $R_*$ at time $t=0$, measured from the beginning of
the decay (the X--ray peak). Then the diffusion equation determining
the disc surface density $\Sigma(R, t)$
\begin{equation}
{\partial\Sigma\over\partial t} = {3\over R}{\partial\over \partial R}
\biggl\{R^{1/2}{\partial\over \partial R}[\nu R^{1/2}\Sigma]\biggr\}
\label{eq5}
\end{equation}
(e.g. Frank et al., 1992) can be rewritten as
\begin{equation}
{\partial\over\partial t}(R^{3/2}\Sigma) 
  = 
{3\nu_*\over R_*}\biggl({\dot M_c\over \dot M_*}\biggr)^{1/4}
    \biggl(R^{1/2}{\partial\over \partial R}\biggr)^2(R^{3/2}\Sigma).
\label{eq6}
\end{equation}
Now defining $p = 2R^{1/2}$ and $S=R^{3/2}\Sigma$ we get finally
\begin{equation}
{\partial S\over \partial t} = 
  {3\nu_*\over R_*}\biggl({\dot M_c\over \dot M_*}\biggr)^{1/4}
  {\partial^2 S\over \partial p^2}.
\label{eq7}
\end{equation}
Thus we obtain a diffusion equation for $S(R,t)$, which is however
nonlinear because of the factor $\dot M_c^{1/4}$ on the rhs: this
itself depends on $S$ and thus $t$ through the relation
\begin{equation}
\dot M_c = 3\pi\nu(R_*,t)\Sigma(R_*, t)
\label{eq8}
\end{equation}
(e.g. Frank et al., 1992). Using (\ref{eq4}) and the definition of $S$
we have
\begin{equation}
\dot M_c = 3\pi\nu_*
\biggl({\dot M_c\over \dot M_*}\biggr)^{1/4}R_*^{-3/2}S(R_*,t),
\label{eq9})
\end{equation}
and thus
\begin{equation}
\biggl({\dot M_c\over \dot M_*}\biggr)^{1/4} =
\biggl[{3\pi\nu_*S(R_*,t)\over R_*^{3/2}\dot M_*}\biggr]^{1/3}.
\label{eq10}
\end{equation}
It is clear that (\ref{eq7}) has a steady solution of the form 
$S = ap + b$, or
\begin{equation}
\Sigma = {a\over R} + {b\over R^{3/2}}.
\label{eq10a}
\end{equation}
Since from (\ref{eq4}) $\nu \propto R$, this is simply the usual steady
solution $\nu\Sigma = A + BR^{-1/2}$ (e.g. Frank et al., 1992), in
which the first term represents a steady inflow with accretion rate
given by $A$, and the second term gives zero net mass inflow.

I now seek separated time--dependent solutions of (\ref{eq7}) of the form
\begin{equation}
S(R,t) = R_*^{3/2}\theta(t)P(p)
\label{eq11}
\end{equation}
with $\theta(t)$ dimensionless and $P$ with the dimensions of
$\Sigma$. Eq. (\ref{eq10}) gives
\begin{equation}
\biggl({\dot M_c\over \dot M_*}\biggr)^{1/4} =
\biggl[{3\pi\nu_*P(p_*)\over \dot M_*}\biggr]^{1/3}\theta^{1/3}(t),
\label{eq12}
\end{equation}
with $p_* = 2R_*^{1/2}$.
Choosing $\theta(0) = 1, P(p_*) = \Sigma(R_*, 0)$, the definition of
$\dot M_*$ implies that the square bracket = 1, so that
\begin{equation}
\biggl({\dot M_c\over \dot M_*}\biggr)^{1/4} = \theta^{1/3}(t).
\label{eq13}
\end{equation}
Then inserting (\ref{eq11}) into (\ref{eq7}) and using (\ref{eq13}) gives
\begin{equation}
{\dot\theta\over 3\theta^{4/3}} = {\nu_*\over R_*}{P''\over P} \equiv
-{1\over T},
\label{eq14}
\end{equation}
where $T$ is a separation constant with dimensions of time. Since
$\theta(0) = 1$ we find
\begin{equation}
\theta = \biggl(1 + {t\over T}\biggr)^{-3}
\label{eq15}
\end{equation}
and also
\begin{equation}
P = A\cos\biggl({R_*\over \nu_* T}\biggr)^{1/2}p +
B\sin\biggl({R_*\over \nu_* T}\biggr)^{1/2}p.
\label{eq16}
\end{equation}
Replacing $p$ by $2R^{1/2}$ and assuming that the disc is 
irradiated out to a fixed outer edge $R_0$, i.e. 
\begin{equation}
\Sigma(R_0,t)=0, 
\label{eq16a}
\end{equation}
we get the eigenvalues
\begin{equation}
T_n = {4R_*R_0\over \nu_*(n-1/2)^2\pi^2},\ \ n=1,2,3,...
\label{eq17}
\end{equation}
and 
\begin{equation}
T'_n = {4R_*R_0\over \nu_*n^2\pi^2},\ \ n=1,2,3,...
\label{eq18}
\end{equation}
for the cosine and sine terms respectively. We can simplify these by
noting from (\ref{eq4}) that
\begin{equation}
\nu_1 \equiv \nu(R_0,0) = \nu_*{R_0\over R_*}, 
\label{eq19}
\end{equation}
so that
\begin{equation}
T_n = {t_{\rm visc}\over (2n-1)^2}, 
\label{eq20}
\end{equation}
\begin{equation}
T'_n  = {t_{\rm visc}\over (2n)^2}
\label{eq21}
\end{equation}
for the two cases, with $t_{\rm visc} = 16R_0^2/\pi^2\nu_1$ a
measure of the viscous
time at $R_0$. Then in terms of the variable 
\begin{equation}
x = {\pi\over 2}\biggl({R\over R_0}\biggr)^{1/2},
\label{eq21a}
\end{equation}
we finally get the general solution of the diffusion
equation for a fully irradiated disc obeying (\ref{eq16a}) as:
\begin{eqnarray}
\lefteqn{\Sigma(R,t) =} \nonumber \\
\lefteqn{{1\over x^3}
\sum_{n=1}^{\infty}\biggl\{{A_n\cos (2n-1)x\over [1+{(2n-1)^2t/ 
t_{\rm visc}}]^{3}}
+{B_n\sin 2nx\over [1+{(2n)^2t/ 
t_{\rm visc}}]^{3}}\biggr\}.}
\label{eq22}
\end{eqnarray}
 
Here the coefficients $A_n, B_n$ are to be found by expanding the
initial surface density distribution $\Sigma(R, 0)$ as a Fourier
series in $\cos (2n-1)x$ and $\sin 2nx$, viz
\begin{equation}
A_n = {4\over \pi}\int_0^{\pi/2}x^3\Sigma (R, 0)\cos(2n-1)x {\rm d}x
\label{eq22a}
\end{equation}
\begin{equation}
B_n = {4\over \pi}\int_0^{\pi/2}x^3\Sigma (R, 0)\sin2nx {\rm d}x.
\label{eq22b}
\end{equation}
(If we had not used the boundary condition (\ref{eq16a}) we should
have found a Fourier transform rather than a series.) For an initial
density distribution with no discontinuities we therefore have $A_n,
B_n \sim 1/n^2$.
Note that the steady solution (\ref{eq10a}) does not obey the boundary
condition (\ref{eq16a}), as mass has to be fed in from $R > R_0$ to
maintain the steady state. Accordingly if we start with initial
conditions $\Sigma (R, 0) = ax^{-2} + bx^{-3}$ in (\ref{eq22}) we do
{\it not} obtain the steady solution, since the expansion assumes
that (\ref{eq10a}) holds. It therefore demands a discontinuity in
$\Sigma(R, 0)$ at $R=R_0$ (i.e. $A_n, B_n \sim 1/n$)
which subsequently diffuses away.

Given the solution (\ref{eq22}), the central accretion rate $\dot
M_c(t)$ can be found from
(\ref{eq12}). (I do not consider here the possible glitch in
the light curve caused by irradiation of the outer disc (cf KR).)
The disc mass within a radius given
by $x_R$ at any time can be expressed as
\begin{equation}
M(R, t) = {64R_0^2\over \pi^3}\int_0^{x_R}x^3\Sigma(x, t){\rm d}x.
\label{eq22c}
\end{equation}
Using (\ref{eq22}) with $x=\pi/2$, the total mass in the $n$th eigenmode is
\begin{eqnarray}
\lefteqn{M_n =
{64R_0^2(-1)^{n+1}A_n\over 
(2n-1)\pi^3[1+{(2n-1)^2t/t_{\rm visc}}]^{3}}} \nonumber \\
\lefteqn{+{64R_0^2((-1)^{n+1}+1)B_n\over 
(2n)\pi^3[1+{(2n)^2t/t_{\rm visc}}]^{3}}.}
\label{eq22d}
\end{eqnarray}
The structure of the expansion (\ref{eq22}) is simple to
understand. The viscous time of the $n$th disc eigenmode varies as 
$t_n \sim t_{\rm visc}/n^2$, reflecting the fact that steeper spatial
gradients cause more rapid diffusion. While the full spectrum of
eigenmodes is in general present in the surface density
distribution at the beginning of the outburst, the higher eigenmodes
will decay more rapidly (as $(1+t/t_n)^{-3}$) as the outburst
proceeds, causing $\Sigma$ to approach the first term of the
expansion ($\propto \cos x$). The disc mass is dominated by this mode in any
case, as from (\ref{eq22d}) we have $M_n \sim n^{-3}(1+t/t_n)^{-3}$. Thus
\begin{equation}
\Sigma(R, t) \sim A_1
\biggl[1+{t\over t_{\rm visc}}\biggr]^{-3} 
\biggl({R_0\over R}\biggr)^{3/2}
\cos{\pi\over 2}\biggl({R\over R_0}\biggr)^{1/2}.
\label{eq23}
\end{equation}
For this term, (\ref{eq12}) gives simply
\begin{equation}
\dot M_c = \dot M_*
\biggl[1+{t\over t_{\rm visc}}\biggr]^{-4}.
\label{eq24}
\end{equation}
Accordingly the X--ray luminosity must decay at least as steeply as this.
From (\ref{eq24}), the typical `exponential' decline phase of SXT
light curves, during which the flux drops by a factor $\sim 10$, corresponds
to $t$ running from 0 to $1.26R_0^2/\nu_1$. The paper by KR derived
the simple law $\dot M_c \propto e^{-t/\tau}$ for the typical
`exponential' decline phase of SXT light curves, with
$\tau = R_0^2/3\nu_1$. The best--fitting function of the form
(\ref{eq24}) to this exponential, over a dynamic range of $10^3$, has
normalization 1.05 and $t_{\rm visc} = 3.11\tau =
1.035R_0^2/\nu_1$, as opposed to $t_{\rm visc} = 16R_0^2/\pi^2\nu_1 =
1.62R_0^2/\nu_1$. Thus the simple treatment of KR agrees surprisingly
well with the present detailed calculation. 
As can be seen from Fig. 1, the power--law decay is almost
indistinguishable from the exponential over this range.
It is much steeper than the $\dot M_c \sim
t^{-(1+\epsilon)}$ decay of an unirradiated disc (see Section 4). 

I note finally in this Section that the fundamental disc outburst mode
(\ref{eq23}) has the mass concentrated quite close to the centre: 
from (\ref{eq22c}) we find 
\begin{equation}
M_1(R, t) = {64R_0^2A_1\over \pi^3[1+t/t_{\rm visc}]^{3}}\sin{\pi\over 2}
\biggl({R\over R_0}\biggr)^{1/2}.
\label{eq24a}
\end{equation}
Thus one--half of the mass of this mode (i.e. effectively one--half of
the total disc mass) lies within $R=0.11R_0$. This
contrasts with outbursts of unirradiated discs, where the usual
$\Sigma \sim R^{-3/4}$ distribution leads to $M(R, t) \sim R^{5/4}$,
so that much of the mass is still in the outer parts of the disc. The reason
for the difference is the larger viscosity in the outer parts of
an irradiated disc (see Section 4 below), which drives the mass
inwards more rapidly. 

This difference in mass distribution holds only in outburst. 
In quiescence irradiation is of course
completely negligible. The disc will evolve in the usual way to a
state in which most of the mass is near its outer edge (i.e. $\Sigma
\sim R$) and begin the next outburst in this state (cf Cannizzo, 1993).

\section{THE LATE DECLINE}
The solutions derived above assume that the whole disc is dominated by
irradiation. However as explained by KR, at late times this cannot be
true, and the region where irradiation is important will retreat
inwards as $R \propto \dot M_c^{1/2}$. Indeed sufficiently large discs
will already be in this regime at the start of the outburst, as even
the Eddington limit luminosity will be unable to ionize hydrogen out
to the edge of such a disc. KR showed that the decline of $\dot M_c$
changes from approximately exponential to linear in this regime. 

The variables $(R, t)$ or $(p, t)$ used in the diffusion equation up to
now are not useful in this regime, as one wants to set boundary
conditions where the disc structure changes from irradiated to
effectively unirradiated, i.e. at a time--dependent radius 
$R \propto \dot M_c^{1/2}$. Accordingly we change the variables from
$(p, t)$ to $(u, t)$, with 
\begin{equation}
u = {p\over \dot M_c^{1/4}(t)}
\label{eq25}
\end{equation}
After some manipulation, the diffusion equation (\ref{eq7}) can then be
cast in the form
\begin{equation}
{\partial S\over \partial t} 
- {u\ddot M_c\over 4\dot M_c}{\partial S\over \partial u}
={3\nu_*\over R_*\dot M_*}
\biggl[{1\over u\dot M_c^{1/4}}{\partial S\over \partial u}
+{1\over \dot M_c^{1/4}}{\partial^2 S\over \partial u^2}\biggr]
\label{eq26}
\end{equation}
One can again look for separated solutions, this time of the form
\begin{equation}
S = U(u)\phi(t),
\label{eq27}
\end{equation}
which implies
\begin{equation}
\dot M_c^{1/4}{\dot\phi\over \phi} = 
{\ddot M_c\over 4\dot M_c^{3/4}}{uU'\over U}
+{3\nu_*\over R_*\dot M_*^{1/4}}\biggl[{1\over u}{U'\over U} +
{U''\over U}\biggr]
\label{eq28}
\end{equation}
Evidently the only hope of finding separated solutions is to require
\begin{equation}
{\ddot M_c\over 4\dot M_c^{3/4}} = A
\label{eq29}
\end{equation}
with $A$ a constant. Then
\begin{equation}
\dot M_c = (At+B)^4
\label{eq30}
\end{equation}
with $B$ also constant. Substituting in (\ref{eq28}) gives
\begin{equation}
(At+b){\dot \phi\over \phi} = Au{U'\over U}
+{3\nu_*\over R_*\dot M_*^{1/4}}\biggl[{1\over u}{U'\over U} +
{U''\over U}\biggr],
\label{eq31}
\end{equation}
which requires each side to equal a separation constant, say
$\beta$. Thus we find
\begin{equation}
\phi = \phi_0(At+B)^{\beta/A}.
\label{eq32}
\end{equation}
Now (\ref{eq10}) and (\ref{eq30}) require $\phi \propto \dot M_c^{3/4} \propto 
(At+B)^3$, which is compatible with (\ref{eq32}) only if
$\beta = 3A$. The $u$ -- equation can be rearranged as
\begin{equation}
U''+ {1\over u}U' + CuU' - 3CU=0,
\label{eq33}
\end{equation}
where $C = AR_*\dot M_*^{1/4}/3\nu_*$. This has a regular singular
point at $u=0$ and an irregular singular point at $u = \infty$,
suggesting a relation to the confluent hypergeometric equation. A
change of independent variable shows that  
\begin{equation}
U = M\biggl(-{3\over 2}, 1; -{Cu^2\over 2}\biggr),
\label{eq34}
\end{equation}
where $M(a, b: u)$ is the confluent hypergeometric (Kummer) function.
The full solution is 
\begin{equation}
\Sigma(u, t) = 
\int f(C)(Ct+D)^3M\biggl(-{3\over 2}, 1; -{Cu^2\over 2}\biggr){\rm d}C,
\label{eq35}
\end{equation}
recalling that $C$ is proportional to $A$. We can use the freedom in
$f(C)$ to satisfy the boundary condition specifying $\Sigma (u_0, t)$
on the irradiation boundary $u=u_0$ where $T_{\rm irr} = T_{\rm
H}$. 

Although (\ref{eq35}) appears complex, its meaning is again
straightforward, and similar to that of the expansion
(\ref{eq22}). The time development $\dot M_c \propto
(At+B)^4$ shows that $A^{-1}$ is a viscous timescale. Since $\dot M_c$
is decaying, we have approximately 
\begin{equation}
\dot M_c \propto \biggl[1 - {t\over t'_{\rm visc}}\biggr]^4,
\label{eq36}
\end{equation}
where $t'_{\rm visc}$ is the longest viscous time of the
partially--irradiated disc. This expression predicts that the outburst
of the irradiated disc shuts off completely at $t = t'_{\rm visc}$. In
reality the unirradiated disc outside the irradiation boundary $u=u_0$
may produce further outburst activity.
Note also that 
the innermost part of the disc may not be irradiation--dominated, as for
small $R$ the effective temperature $T_{\rm eff}$ is larger
than $T_{\rm irr}$ for similar local accretion rates.

Equation (\ref{eq36}) thus gives an approximate description of the
regime in which the central emission is unable to keep the disc
ionized all the way to its outer edge. KR found a linear decay of the
central X--rays in this case, which is approximately what is
observed (Giles et al., 1996) in the decline of GRO J1744--28, the
transient with the longest known orbital period. As can be seen
(Fig. 2), eq. (\ref{eq36}) gives a rather similar light curve over the
typical observed dynamic range. Note in particular that the linear
decay of GRO J1744-28 extends only over a dynamic range of about 4 in
flux. After this the observed flux is persistently higher than the
linear trend, just as in Fig. 2.

\section{DISCUSSION}

I have shown that analytic solutions for the viscous evolution of an
irradiated disc predict a steep power--law decay in the central
accretion rate $\dot M_c$. In practice this is quite close to the exponential
decay predicted 
by the dimensional arguments of KR; this is a good representation of typical
observed X--ray light curves. It is important to ask why these
solutions are so different from the very shallow $\dot M_c \sim
t^{-(1+\epsilon)}$ decay of an unirradiated disc. 

As pointed out by
Cannizzo (1997), the latter is easy to understand. The equation for
vertical radiative energy transport within a disc
\begin{equation}
{4\sigma T_m^4\over 3\kappa\Sigma} = {9\over 8}\nu\Sigma {GM\over R^3}
\label{eq37}
\end{equation}
(where $T_m$ is the midplane temperature)
can be combined with the alpha--prescription (\ref{eq1}) to give
\begin{equation}
\nu \propto \kappa^{1/3}\Sigma^{2/3}R,
\label{eq38}
\end{equation}
where $\kappa$ is the opacity. The factor $\kappa^{1/3}$
generally varies only slowly with $\Sigma$ and
$T_m$, so the diffusion equation (\ref{eq5}) predicts
\begin{equation}
{\partial\Sigma\over\partial t} \propto \Sigma^{5/3} 
\label{eq39}
\end{equation}
and thus $\Sigma \sim t^{-3/2}$ 
for a separated solution for an unirradiated disc.

The fundamental difference for an irradiated disc is that, from (\ref{eq4}),
the viscosity $\nu$ is {\it independent} of the local value of $\Sigma$.  Thus
(\ref{eq39}) is replaced by
\begin{equation}
{\partial\Sigma\over\partial t} \propto \dot M_c^{1/4}\Sigma. 
\label{eq40}
\end{equation}
Were it not for the weak $\dot M_c^{1/4}$ dependence we would
get an exactly exponential decay for $\Sigma(t)$ and thus for $L_X$: this is
in any case clearly a good approximation until  $\dot M_c$ has
decreased by large factors. Note that
this dependence of $\nu$ on disc properties is {\it global}, i.e. on
the value of $\dot M$ or $\Sigma$ at the inner edge of the disc, rather
than a local dependence on $\Sigma$ at each radius. The effect is to
convert the
exponentials into steep power laws since the relation $\dot M_c
\propto \Sigma(R_*)^{4/3}$ implies $\ddot M_c \propto \dot M_c^{5/4}$
and thus $\dot M_c\sim t^{-4}$ for separated solutions. 

The different viscosity behaviour in an irradiated disc is also responsible
for the greater central concentration of the disc mass
noted at the end of Section 2. The unirradiated viscosity (\ref{eq38})
drops as $\Sigma$ decreases, tending to slow the depletion of the
lower surface density of the outer disc regions. By contrast the
irradiated viscosity (\ref{eq4}) is independent of the local
value of $\Sigma$ and increases with $R$, thus driving mass inwards
from large radii. The depleted density in the outer regions implies a
smaller local accretion rate there: from (\ref{eq4}) and the local
equivalent of (\ref{eq8}) we find
\begin{equation}
\dot M(R, t) = 
\dot M_c(t)\biggl({R_*\over R}\biggr)^{1/2}{\cos{\pi\over 2}(R/R_0)^{1/2}
\over \cos{\pi\over 2}(R_*/R_0)^{1/2}},
\label{eq41}
\end{equation}
so that the local effective temperature goes as $\sim R^{-7/8}$
instead of the usual $\sim R^{-3/4}$.

\section{CONCLUSIONS}

The presence of a central irradiating point source completely changes
the viscous evolution of an accretion disc. Because the viscosity is
controlled by conditions in the centre, the disc mass in outburst
is more centrally
condensed than for unirradiated discs. Most importantly, the decay of
the central accretion rate is much steeper, being approximately
exponential. I conclude that irradiated discs give a good
representation of soft X--ray transient outbursts.

\section{ACKNOWLEDGMENTS}

I thank Hans Ritter, John Cannizzo and Henk Spruit for very helpful
discussions, Graham Wynn for help with curve--fitting, and the
U.K. Particle Physics and Astronomy 
Research Council for a Senior Fellowship. Theoretical astrophysics
research at Leicester is supported by a PPARC Rolling Grant.

\noindent {\bf FIGURE CAPTIONS} \\
\newline
\noindent
{\bf Figure 1.} Comparison of the exponential light curve $L_x \propto
e^{-t/\tau}$ (solid curve) derived by KR, and the best--fitting curve
of the form (\ref{eq24}) of this paper 
(dashed curve): this has $L_x \propto 1.05(1+t/(3.11)\tau)^{-4}$. 
\newline
\noindent
{\bf Figure 2.} Comparison of the linear light curve $L_x \propto 1 -
t/\tau'$ derived by KR (solid curve), and the best--fitting curve of
the form (\ref{eq36}) of this paper (dashed curve): this has 
$L_x \propto 1.073(1 - t/2.647\tau')^4$.

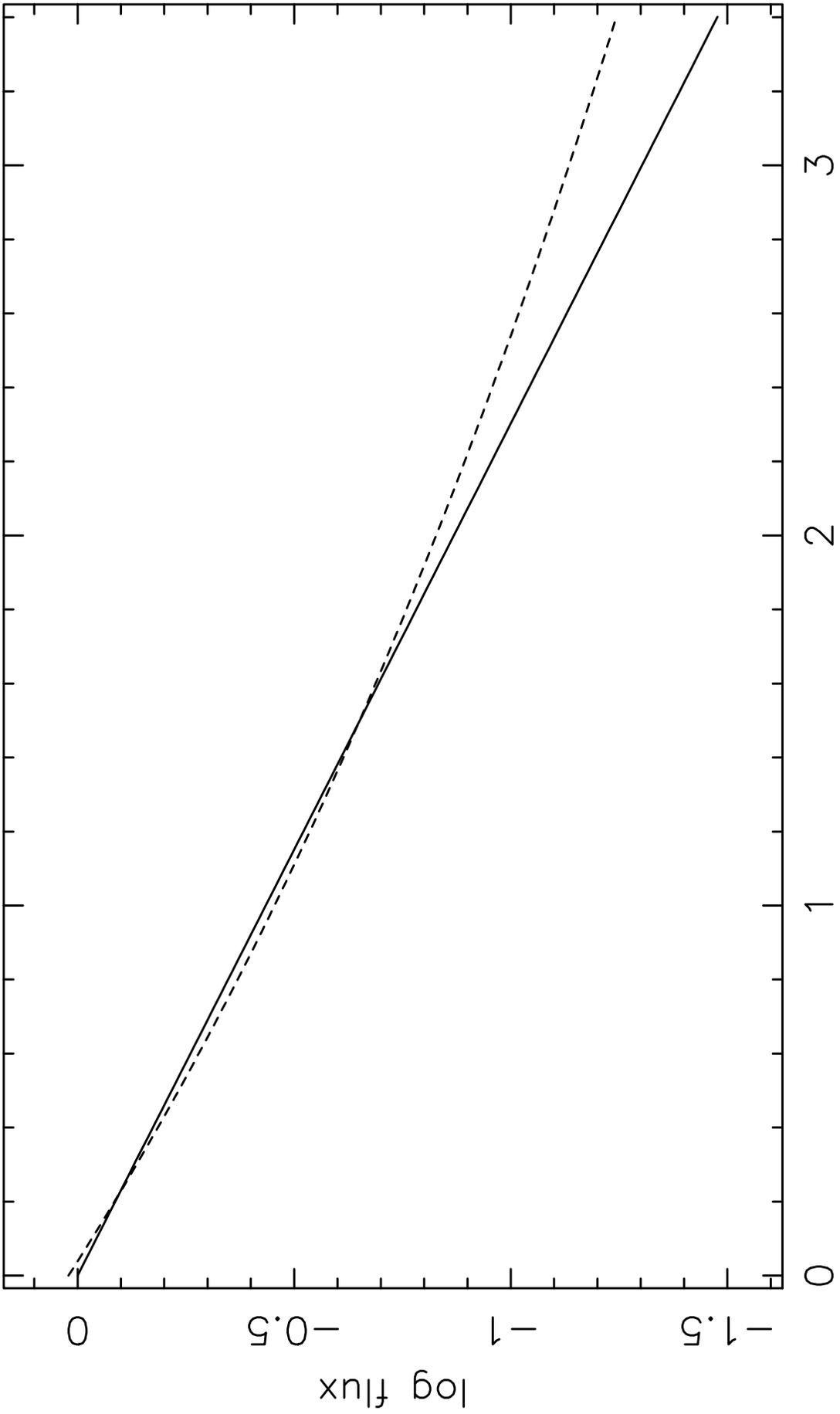
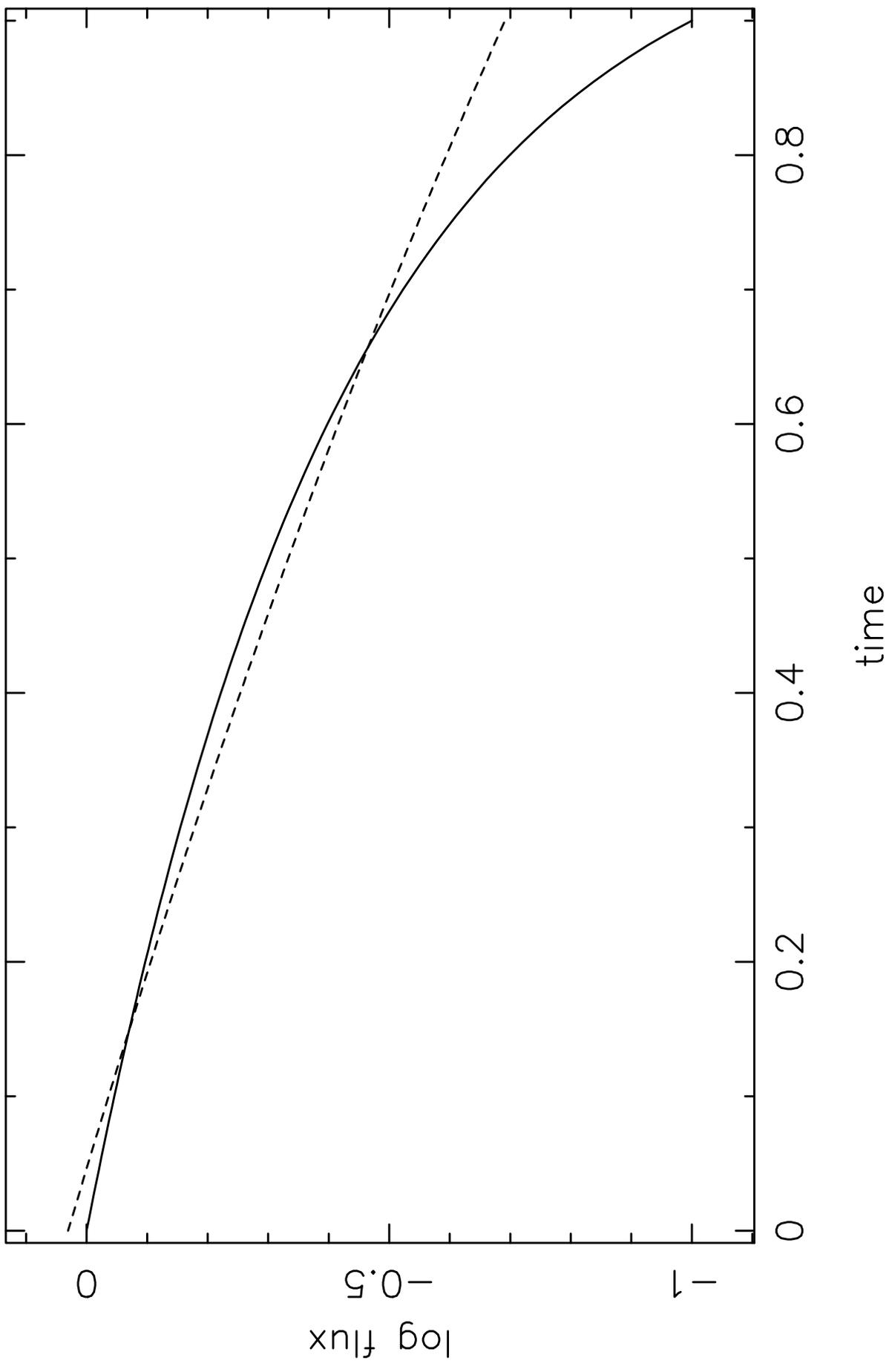
\end{document}